\documentclass[sigconf,screen]{acmart}

\setcopyright{acmlicensed}
\acmPrice{15.00}
\acmDOI{10.1145/3611643.3613860}
\acmYear{2023}
\copyrightyear{2023}
\acmSubmissionID{fse23industry-p5-p}
\acmISBN{979-8-4007-0327-0/23/12}
\acmConference[ESEC/FSE '23]{Proceedings of the 31st ACM Joint European Software Engineering Conference and Symposium on the Foundations of Software Engineering}{December 3--9, 2023}{San Francisco, CA, USA}
\acmBooktitle{Proceedings of the 31st ACM Joint European Software Engineering Conference and Symposium on the Foundations of Software Engineering (ESEC/FSE '23), December 3--9, 2023, San Francisco, CA, USA}

\usepackage{xspace}
\usepackage{listings}
\usepackage{capt-of}
\usepackage{flushend}
\usepackage{subfig}
\usepackage{multirow}
\usepackage{afterpage}
\usepackage{placeins}
\usepackage{multicol}
\usepackage{adjustbox}
\usepackage{tabularx}

\makeatletter
\def\namedlabel#1#2{\begingroup
\def\@currentlabel{#2}%
\phantomsection\label{#1}\endgroup
}
\makeatother

\newcommand{\chref}[1]{\ref{#1}}

\newcommand{\INCA}{IncA\xspace}
\newcommand{\IQL}{iQL\xspace}
\newcommand{\QL}{QL\xspace}
\newcommand{\CODEQL}{CodeQL\xspace}
\newcommand{\VQ}{\textsc{VQ}\xspace}
\newcommand{\VQL}{\textsc{VQL}\xspace}
\newcommand{\SM}{semi-na\"ive\xspace}

\newcommand{\ic}[1]{\changefont{cmtt}{m}{n}{#1}\normalfont}  
\newcommand{\changefont}[3]{\fontfamily{#1}\fontseries{#2}\fontshape{#3}\selectfont}
\newcommand\parhead[1]{\vspace{1mm}\noindent\textbf{{#1}}\ \ }

\definecolor{lightlightgray}{gray}{0.9}
\definecolor{codecolor}{gray}{0.15}
\definecolor{javadocblue}{rgb}{0,0,0.8} 
\definecolor{blackblack}{rgb}{0,0,0}

\lstdefinestyle{numbers} {numbers=right, numberstyle=\tiny,stepnumber=1,numbersep=5pt}

\begin{document}

\title{Incrementalizing Production CodeQL Analyses}

\author{Tamás Szabó}
\affiliation{
\institution{GitHub}
\country{Germany}
}

\begin{abstract}
Instead of repeatedly re-analyzing from scratch, an incremental static analysis only analyzes a
codebase once completely, and then it updates the previous results based on the code changes.
While this sounds promising to achieve speed-ups, the reality is that sophisticated static
analyses typically employ features that can ruin incremental
performance, such as inter-procedurality or context-sensitivity.
In this study, we set out to explore whether incrementalization can help to achieve speed-ups for
production \CODEQL analyses that provide automated feedback on pull requests on GitHub.
We first empirically validate the idea by measuring the potential for reuse on real-world
codebases, and then we create a prototype incremental solver for \CODEQL that exploits
incrementality.
We report on experimental results showing that we can indeed achieve update times proportional
to the size of the code change, and we also discuss the limitations of our prototype.
\end{abstract}



\maketitle

\section{Introduction}
\label{sec:Introduction}

Static analyses play a key role in modern software development because they help catch potential
runtime errors already at development time.
\CODEQL~\cite{CODEQL} is a static analysis framework that has seen widespread adoption in industry.
\CODEQL analyses are written in a declarative language called \QL, which compiles to Datalog under the hood.
The subject program (i.e.\ codebase to be analyzed) is extracted into a relational database
format, and the \CODEQL solver executes analyses as queries over the database.
GitHub Code Scanning comes with a suite of \CODEQL analyses for various languages that provide automated
feedback about security vulnerabilities and potential runtime issues at pull request (PR) review
time.

Performance and precision are key requirements when it comes to static analyses.
On the one hand, the execution time is a key factor for the adoption of static analyses.
Prior research shows that static analyses have at most a few minutes to compute their results when
they are used as part of a code review process in continuous-integration~(CI)
environments~\cite{WHY_DONT_DEVS_USE_ANALYSES,WHAT_DEVS_WANT_AND_NEED,TRICORDER}.
Failing to meet this timing constraint interrupts the development flow, which quickly leads to
disuse of the analyses.
On the other hand, developers expect that static analyses faithfully capture the program behavior,
so that they do not report errors that cannot actually occur at runtime.
Unfortunately, these two requirements are at odds with each other, and static analysis designers
must carefully strike a balance between them.
\CODEQL approaches this trade-off from the angle of precision because \CODEQL analyses are highly
sophisticated by design.
\CODEQL supports many popular programming languages, and, for all these languages, the \CODEQL
analyses use features and building blocks that are responsible for good precision;
ranging from inter-procedural reasoning through context-sensitivity to field-sensitivity.
This also means that the \CODEQL solver has a difficult task, which it addresses by using
a range of optimizations developed in the Datalog and database community~\cite{CODEQL,DatalogBible}.
Still, for large projects, even this is not enough, and the execution time can grow well beyond
the desired few minutes ballpark.

Researchers have long been investigating various techniques that can help speed up static
analyses.
Examples include parallel execution~\cite{PARALLEL_POINTSTO,CAULIFLOWER}, compositional
analysis formulation~\cite{JIT_ANALYSIS,FACEBOOK_INFER}, partial evaluation~\cite{I3QL14,SOUFFLE}, demand-driven execution~\cite{BOOMERANG,INC_PT_DATALOG}, or incrementalization~\cite{TAMAS_THESIS}.
When it comes to delivering updated analysis results on a PR,
incrementalization seems like an obvious choice.
This is because, instead of repeated re-analysis from scratch, an incremental analysis reuses the
previously computed analysis results and updates that based on the changed code parts.
This aligns with the life cycle of a PR because that includes multiple rounds of reviews with
small code changes in each iteration.
If the computational effort required to update the previous results is proportional to the size
of the code change (and not the size of the entire subject program), then significant speed-ups \emph{can}
be achieved.
Prior work shows that this kind of speed-up is a reality for real-world static
analyses~\cite{INCA_PLDI,INC_PT_DATALOG,COCO,I3QL14}, but \CODEQL does not employ
incrementalization yet.
It is important to emphasize that it is not necessarily the case that a small input
change \emph{always} leads to a small change in the analysis result.
All those features that make static analyses precise (e.g.\ inter-procedurality, context-sensitivity)
directly go against the potential benefits of incrementalization because even a
small program change can have far-reaching transitive effects requiring re-analysis of a
significant part of the subject program.
There is prior work on investigating the incrementalizability of analyses
implemented in Datalog~\cite{INCA_PLDI,ELASTIC_SOUFFLE}, but the results are inconclusive, and
they do not obviously generalize to \CODEQL.

In this paper, we conduct a \emph{study} to investigate if and how we could use incrementalization
to speed up production \CODEQL analyses.
We start off by clearing our doubts about whether incrementalization can be effective at all for
\CODEQL analyses.
Following the idea of \citeauthor{INCA_PLDI}~\cite{INCA_PLDI}, we conduct an \emph{impact
benchmark} using real-world Ruby codebases with a sophisticated data flow analysis.
Our goal is to empirically reason about how much of the previously computed analysis results can
be reused on average throughout a series of real-world commits, thereby measuring how much
potential benefit we could get from an incremental analysis.
We conduct this experiment without an actual incremental solver, just by using the existing \CODEQL
analysis pipeline.
We find that there is a lot of potential for reuse and, with that, for speed-ups in updating
previously computed results.
This observation proves for us that it indeed makes sense to experiment with prototyping an
incremental solver for \CODEQL.

Given the complexity of the existing \CODEQL solver, it is a daunting task to rewrite that to be
incremental.
Instead, for the purpose of our study, we prototype an incremental solver termed \IQL
based on an off-the-shelf incremental Datalog solver called Viatra Queries (\VQ)~\cite{VIATRA}.
We chose \VQ because it has been reported to deliver good incremental performance and because it
is already used as the back end of an incremental static analysis framework called \INCA~\cite{INCA_ASE}.
The challenge in our work is that while \VQ has good expressive power in terms of the kinds of
operations it can incrementalize, it is still not sufficient when compared to the kinds of
analyses \QL can express.
We show how we transform the \QL representation of an analysis to the representation that \VQ can
understand while bridging the abstraction gap between the two systems.
Given that \CODEQL scales significantly better than \VQ in terms of raw performance, we also
show how to execute parts of an analysis non-incrementally with \CODEQL in an otherwise
incremental evaluation governed by \VQ.
Ultimately, we end up with a prototype incremental solver that can execute production \CODEQL
analyses fully incrementally or in a hybrid setting where the non-incremental and incremental
modes are mixed.

We measure the performance of \IQL on real-world Ruby projects by running the above-mentioned
\CODEQL analysis incrementally on their commit histories.
We find that the fully-incremental analysis takes maximum \textasciitilde15~seconds to update analysis results
for commits affecting up to 1000 lines of code.
However, it takes more than an hour to perform the first from-scratch analysis, and the memory use (due to caching)
of the analysis can go as high as \textasciitilde70 GB, which is prohibitive.
This is where the hybrid approach shines because it presents an interesting trade-off opportunity.
We find that by executing all non-recursive parts of the analysis implementation non-incrementally,
we manage to reduce the initialization time to \textasciitilde15 minutes and the memory use to \textasciitilde20 GB\@.
In turn, the incremental update time gets higher compared to the fully-incremental approach, but
it is still below a minute.
In both cases, we find that the incremental update time is actually proportional to the size of
the commit.
These results are promising for \CODEQL because they show that incrementalization can deliver fast
feedback, but we also acknowledge that the high memory use requires further work before
incrementalization can make its way to production \CODEQL.

This paper makes the following contributions:
\begin{itemize}
\item We identify the challenges that come with the incrementalization
of production \CODEQL analyses (\autoref{sec:Problem}).
\item We present our impact benchmark which investigates whether incrementalization can be
beneficial at all for production \CODEQL (\autoref{sec:Impact}).
\item We develop \IQL based on an existing incremental Datalog
solver called \VQ. We discuss how we transform the \QL analysis representation to \VQ, and we
also discuss the idea of the hybrid solver (\autoref{sec:Backend}).
\item We benchmark the performance of \IQL and show that incrementalization can deliver
fast enough feedback for PRs (\autoref{sec:Evaluation}).
\end{itemize}
\section{Challenges and High-level Solution Approach}
\label{sec:Problem}

We present a motivating example that shows \CODEQL in action.
Then we set the stage for our study to figure out whether incrementalization can
help at all to speed up \CODEQL analyses.

\subsection{Motivating Example}
\label{sec:Problem-Example}

\begin{figure}
\centering
\includegraphics[width=\columnwidth]{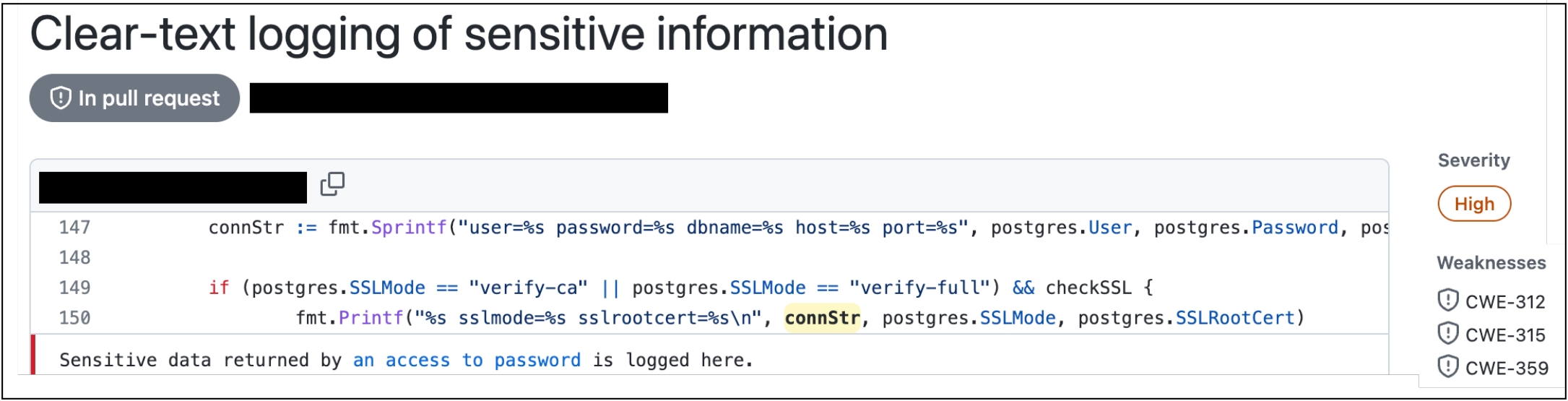}
\caption{A concrete security vulnerability identified by \CODEQL in a Go codebase.}
\vspace{-5mm}
\label{fig:Vulnerability}
\end{figure}

\CODEQL analyses are widely used in production to identify security vulnerabilities in GitHub
repositories.
\autoref{fig:Vulnerability} shows an example where \CODEQL reported a vulnerability in
a Go codebase as part of a PR review.
This vulnerability is a clear-text logging of sensitive information, and it is also
categorized in the Common Weakness Enumeration (CWE) database.\footnote{\url{https://cwe.mitre.org/}}
Based on this feedback, developers can fix the vulnerability before the code gets merged
into the main branch and then shipped to production.
Even though this vulnerability is about a specific analysis and language, \CODEQL supports many
programming languages and frameworks,\footnote{\url{https://codeql.github.com/docs/codeql-overview/supported-languages-and-frameworks/}} and it
comes with a rich set of analyses covering many important vulnerabilities.\footnote{\url{https://codeql.github.com/codeql-query-help/full-cwe/}}

The standard analysis suite for Go contains, among others, the above-mentioned clear-text logging
analysis and takes around 6 minutes to run on this particular codebase comprising 10~KLoC\@.
While the feedback comes as part of a PR review in \autoref{fig:Vulnerability}, the size of the
code change in the PR does not actually matter for the analysis: It executes from scratch on
the whole codebase every time the PR is updated.
For larger projects, the run time can get significantly higher, and it needs to be paid repeatedly.
Research shows that developers are willing to wait at most a few minutes for automated feedback
on a PR, afterwards they switch context, and the usefulness of the feedback quickly
degrades~\cite{WHAT_DEVS_WANT_AND_NEED,WHY_DONT_DEVS_USE_ANALYSES}.
Our goal is to make the analysis time proportional to the size of the code change.
To this end we conduct a study where we experiment with incrementality to deliver fast updates
for PRs.

\vspace{-5mm}
\subsection{Prior Work on the Incrementalizability of Static Analyses}
\label{sec:Problem-Incrementalizability}

Static analyses often use inter-procedurality or context-sensitivity to deliver good precision.
However, these are exactly the features that make an analysis computationally expensive.
They are also the reasons why it is not straightforward to answer whether incrementality can help
to achieve speed-up for such analyses.
For example, if an analysis reasons about the subject program inter-procedurally, then it is not
necessarily true that a small program change always leads to a small change in the analysis
result because the change may have far-reaching effects across transitively reachable functions
in the call graph.
Inter-procedurality is essential to almost all production \CODEQL analyses because
vulnerabilities are rarely local to a single function but instead manifest across a series of
function calls.
Thus, it is a question if incrementality can help at all to speed up \CODEQL analyses.

Prior research has already investigated the incrementalizability of inter-procedural analyses
implemented in Datalog.
For example, \citeauthor{INCA_PLDI} used inter-procedural points-to, interval, and constant
propagation analyses on Java codebases from the Qualitas corpus~\cite{
INCA_PLDI,QUALITAS}.
They measured performance on synthesized IDE-style program changes to simulate the kind of
changes that developers typically make in an IDE, modifying only individual expressions or
statements.
This is in contrast to larger program changes, as one might see in a commit.
The kinds of program changes they introduced are tailored to trigger the worst-case behavior of
the analyses in the sense that, e.g., changing an allocation site will affect the result of a
points-to analysis with high probability, so that the analysis will definitely have work to do to
update the analysis result.
They reason about the \emph{impact} of program changes.
They introduce the notion of impact based on how the analysis result is represented in Datalog,
which is about storing both the subject program and the analysis result as sets of
relations in databases.
They measure impact as the size of the diff between two database snapshots (representing two
analyses results) in terms of the overall number of tuples that get inserted or deleted in any
relation.
They find that the majority of the program changes have small impact, which is a necessary
condition for incrementalizability.

\citeauthor{ELASTIC_SOUFFLE} also investigate this topic with an inter-procedural points-to
analysis adapted from Doop running on a Java codebase~\cite{ELASTIC_SOUFFLE,DOOP}.
They also synthesize program changes: They randomly select a subset of the tuples (up to 1000)
from the database that represents just the subject program (without the analysis result), and
they delete and then re-insert those tuples while measuring the impact as explained before.
They find large variability in impacts and argue that incrementalization does not
necessarily always help.
They create an incremental solver that does not actually always perform an incremental update, but,
based on heuristics, restarts the analysis from scratch if the impact of a change is deemed too
high.
It is not clear if these two pieces of related work contradict each other because they use
different analyses and code changes, but they are indicators that more exploration is needed to
figure out if \CODEQL analyses can benefit from incrementalization with commit-style changes.
This leads us to our first challenge:

\vspace{1mm}
\noindent \fbox{
\parbox{0.95\columnwidth}{
\noindent \textbf{Incrementalizability~(C1):}\namedlabel{ch:incrementalizability}{\textbf{
Incrementalizability~(C1)}}
Does incrementalization have potential for speeding up production \CODEQL analyses?
Can we answer this without actually implementing an incremental solver?
}
}
\vspace{1mm}

\parhead{Solution approach} We also conduct an impact measurement. We use
real-world Ruby codebases with their commit histories and a sophisticated \CODEQL data flow analysis
and show that incrementalization indeed has potential to speed up \CODEQL analyses (\autoref{sec:Impact}).

\subsection{Challenges of Incrementalizing \CODEQL}
\label{sec:Problem-CodeQL}

Assuming that the impact measurement yields promising results, we now want to investigate
what it would take to actually incrementalize the \CODEQL analysis pipeline and the
challenges that come with that. We must first briefly review the main components of \CODEQL.

\begin{figure}
\centering
\includegraphics[width=0.8\columnwidth]{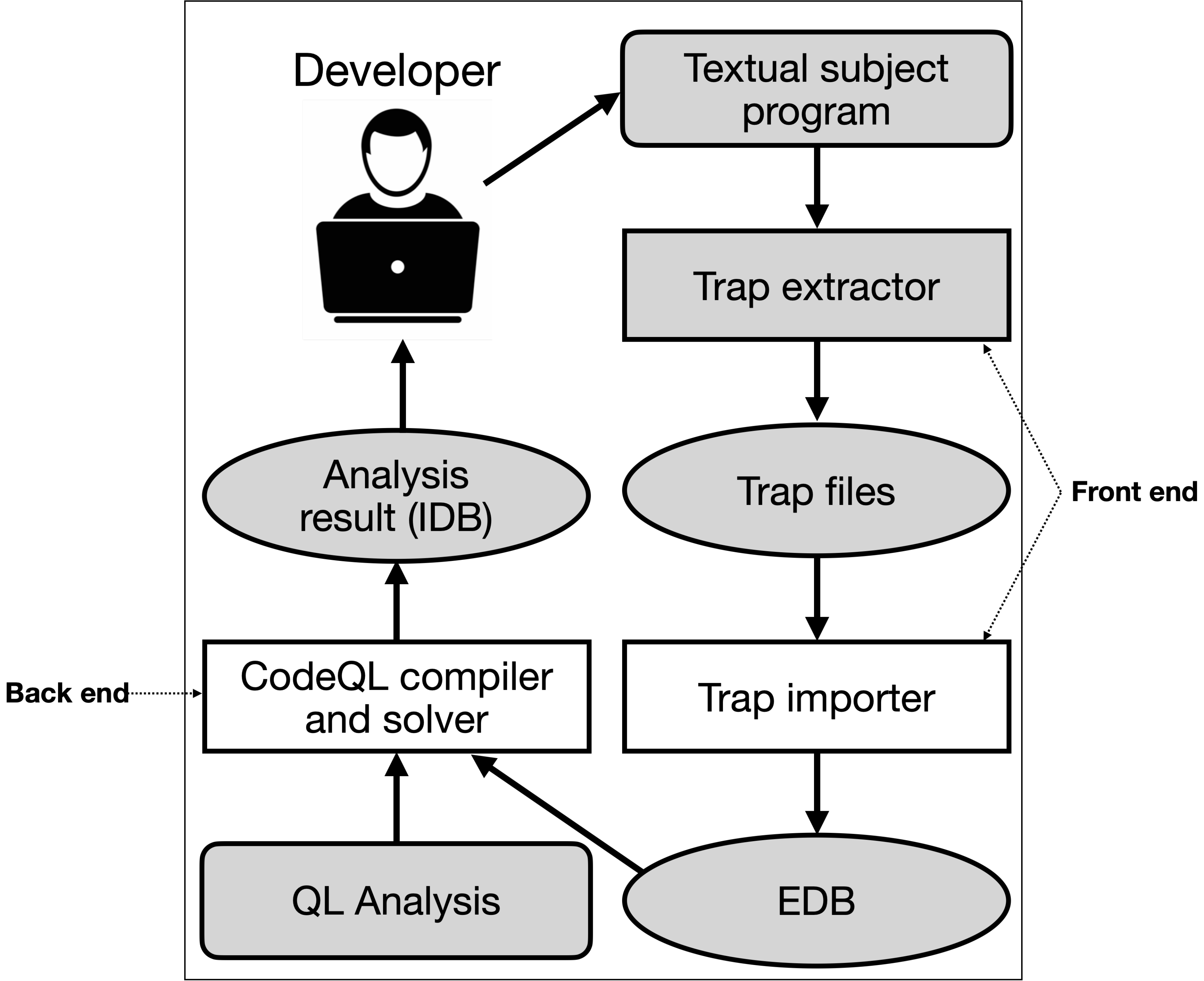}
\caption{High-level architecture of the \CODEQL analysis pipeline. Rectangles are individual analysis
components. Rounded rectangles are the inputs to \CODEQL: the subject program and the analysis.
Ovals are output artifacts. Language-specific components are shaded.}
\label{fig:Architecture}
\end{figure}

\autoref{fig:Architecture} shows the high-level architecture of \CODEQL.
The subject program is given to \CODEQL as text, i.e., the code that is stored in a
GitHub repository.
The \emph{trap extractor} is a language-specific component that creates so called \emph{trap} files.
Trap files encode various information about the subject program that \CODEQL analyses later could
require, such as an abstract syntax tree (AST) representation of the program, represented in
a relational format.
Each supported language has its own database schema defining the available relations.
For dynamically-typed languages, such as Ruby or Python, the trap extractor only extracts
information from the AST\@.
For other languages, the trap extractor may require compilation and build tasks, so that it can
also extract derived information about the program, i.e.\ type hierarchy or macro expansions.
The \emph{trap importer} is language-independent, and it builds the extensional database (EDB)
from a set of trap files.
The EDB represents the subject program in relational form.

The \emph{\CODEQL compiler and solver} takes the EDB and the \QL analysis definition as input to
first compile the analysis and then compute the \emph{analysis result} which is also called the
intensional database (IDB).
The \QL analysis is defined with the declarative \QL language, and the \CODEQL solver completely
hides the execution details.
It is important to emphasize that \CODEQL does not usually store the entire IDB explicitly but
only populates that on demand.
Using the analysis result, we can then provide feedback to developers on the GitHub UI\@.

Incrementalization of the \CODEQL analysis pipeline requires that every component of the pipeline
is either truly incremental or fast enough that it does not become a bottleneck for performance.
This leads us to two more challenges:

\vspace{1mm}
\noindent \fbox{
\parbox{0.95\columnwidth}{

\noindent \textbf{Front end~(C2):}\namedlabel{ch:frontend}{\textbf{Front end~(C2)}} Can we make
the \CODEQL front end (trap extractor and importer) incremental?

\noindent \textbf{Back end~(C3):}\namedlabel{ch:backend}{\textbf{
Back end~(C3)}} Can we create an incremental solver that supports production \CODEQL
analyses?
}
}
\vspace{1mm}

\noindent Finally, we also want to get empirical evidence about performance, which leads us to
the following challenges:

\vspace{1mm}
\noindent \fbox{
\parbox{0.95\columnwidth}{
\noindent \textbf{Update time~(C4.1):}\namedlabel{ch:updateTime}{\textbf{Update time~(C4.1)}}
Can we achieve incremental update times that are (i) below a minute on average per commit and (ii)
proportional to the size of the analyzed commit? The minute threshold is reasonable for PR-style
feedback based on prior studies in this area~\cite{WHAT_DEVS_WANT_AND_NEED, WHY_DONT_DEVS_USE_ANALYSES}.

\noindent \textbf{Init time~(C4.2):}\namedlabel{ch:initTime}{\textbf{Init time~(C4.2)}} Does our
incrementalization approach come with an acceptable initialization time?

\noindent \textbf{Memory use~(C4.3):}\namedlabel{ch:memory}{\textbf{Memory use~(C4.3)}} Is the
extra memory use induced by incrementalization acceptable? \CODEQL analyses are typically executed
on standard GitHub Actions runners with 7~GB of memory,\footnotemark so we ideally want to stay
below that threshold.
}
}\footnotetext{\url{https://gh.io/AAjpnuo}}
\vspace{1mm}

\parhead{Solution approach} When it comes to the front end of \CODEQL, it turns out that
non-incremental parsers are already able to parse source files in milliseconds which is fast enough
to enable their use even in an incremental analysis pipeline~\cite[Chapter~7]{TAMAS_THESIS}.
We acknowledge though that for compiled languages \CODEQL requires more than just parsing, so this
is not a full solution in general, but it is still sufficient for our study given that we target
Ruby.
To this end, we focus on the construction and incremental maintenance of the EDB\@.
We show that the current \CODEQL front end is not compatible with incrementalization because
it is unable to reuse parts of the EDB that belong to unchanged code parts which degrades
incremental performance later in the analysis phase.
The root cause of this issue lies in the way how ids are generated for program elements
in the EDB\@.
We fix this problem by using an alternative id generation approach which ensures that unchanged
program elements get the same ids assigned across commits (\autoref{sec:Impact}).
Next, we create an incremental solver called \IQL for \QL analyses based on an off-the-shelf
incremental Datalog solver called \VQ (\autoref{sec:Backend}).
Finally, we benchmark the performance of \IQL on real-world Ruby projects (\autoref{sec:Evaluation}).
\section{Background on \CODEQL and \VQ}
\label{sec:Background}

Before we delve into the details of our approach, we first present some background material on \CODEQL and \VQ.

\subsection{\CODEQL}
\label{sec:Background-CODEQL}

\parhead{\QL language features}
While the surface syntax of \QL looks quite unlike Datalog, it actually gets compiled
down to a fairly standard Datalog dialect called DIL\@.
Recursion is central to \QL, and it is implemented using
least fixpoint semantics. Since \QL also supports negation and non-monotonic
aggregates, restrictions have to be put into place to ensure a well-defined
semantics. Other Datalog dialects generally insist on strict stratification
whereby recursion through negation is not allowed~\cite{DatalogBible}.
\QL relaxes this requirement and allows parity-stratified negation which
means that recursive calls under an even number of negations are allowed, though
recursion through other non-monotonic operators is still forbidden.%
\footnote{It is important to note here that \QL (and DIL) are not
Horn-clause based, and rule bodies can contain arbitrarily nested
first-order logic formulas. In Horn clauses, of course, calls always appear
under at most one negation, so parity stratification does not make sense.}

\QL also defines a number of \emph{built-ins} that are awkward or impossible to express in
standard Datalog to provide operations that are frequently used in static analyses, such as
computation of dominator trees or shortest paths.
Built-ins have implementations in Java, which \CODEQL invokes during the evaluation process.
\QL also supports user-defined algebraic data types (ADTs), which are used heavily in the
standard analysis suite. This is a big jump in expressive power considering that standard Datalog
does not allow the construction of any kind of new value.
Virtually every \QL analysis uses ADTs. At runtime, ADTs also require the generation of fresh ids
in order to distinguish between the various instances~\cite{ipa}.
Finally, \QL supports complex recursive aggregations (including non-monotonic ones) which are
challenging even in a non-incremental setting~\cite{doing-a-doaitse}.

\begin{figure}
\centering
\includegraphics[width=\columnwidth]{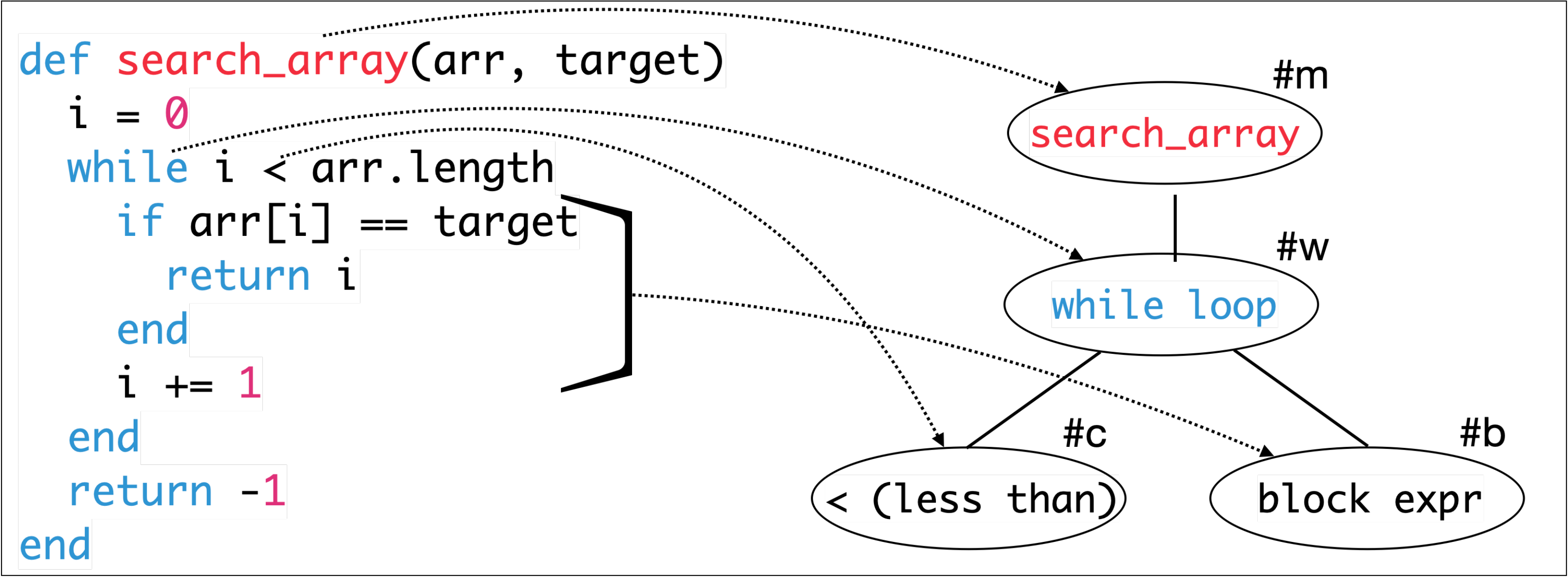}
\caption{An example Ruby code snippet and a fragment of its AST. Each AST node has an associated
label.}
\vspace{-5mm}
\label{fig:AST}
\end{figure}

\parhead{Front end}
We already discussed that both the subject program and the analysis result are represented in
relations.
The trap extractor emits instructions in trap files that will determine what kind of
tuples get inserted into the EDB during trap import.
As an example, consider the code snippet on the left in \autoref{fig:AST} and a fragment of its AST on the
right.
A fragment of the trap file associated with this code snippet looks as follows.
\begin{lstlisting}[frame=lrtb,morekeywords={assert},keepspaces=true,xleftmargin=0mm,xrightmargin=
0mm]{Name}
#m = @"example.rb#search_array"
#w = *, #b = *, #c = *
ruby_method_def(#m, 'search_array(arr, target)')
ruby_while_def(#w, #b, #c)
\end{lstlisting}
The trap extractor uses \emph{keys} to refer to program elements.
Keys will get turned into integer ids later by the trap importer, but the actual values of integer
ids do not matter at extraction time.
Global keys are represented with a string value, and multiple occurrences of the same global key
will get the same id assigned through string interning.
For example, we assign the global key \ic{@"example.rb\#search\_array"} to \emph{label} \ic{\#m}.
Here, we prepend the container file's path before the name of the method because this gives us a
globally unique reference to the method.
For the other AST nodes, we use local keys (\ic{*}), which simply prescribes for the trap
importer the generation of a fresh integer id without sharing of the value.
The trap importer uses a globally incremented counter, and it rolls out ids sequentially.
We access the local keys through the other labels (\ic{\#w}, \ic{\#b}, \ic{\#c}) and then emit
one tuple each in relations \ic{ruby\_method\_def} and \ic{ruby\_while\_def}.
The trap importer will instantiate the real tuples later and insert them into the EDB\@.

\parhead{Back end}
\QL code first gets compiled to DIL, which, in turn, gets compiled to a low-level relational
algebra (RA) representation.
The \CODEQL compiler performs numerous optimizations both on the DIL and
on the RA level, including constant folding, dead-code elimination, and join ordering.
Join ordering is particularly important because real-world codebases typically produce relations
with millions of tuples, so a single inefficient join can make the evaluation intractable.
The evaluation of recursive analyses is challenging in general in Datalog.
\CODEQL employs a well-known technique called \SM evaluation for this purpose.
This means that a recursive program gets evaluated with an iterative fixpoint
computation where each iteration only considers newly computed results from the previous
iteration to improve performance~\cite{DatalogBible}.
At least a conceptual level, \SM evaluation operates with different
auxiliary predicates that compute e.g.\ the set of all tuples computed in the previous iteration,
the set of new tuples (i.e.\ not inferred before) computed in the previous iteration, or the
tuples computed thus far in the current iteration.
Typically, Datalog solvers perform \SM computation implicitly without requiring the
auxiliary predicates to be present in the input Datalog program.
However, \CODEQL generates the helper predicates explicitly during compilation, and the
presence of the predicates is heavily woven into the downstream compilation stages.

\subsection{Viatra Queries}
\label{sec:Background-VQ}

\VQ is a model transformation framework primarily used for Eclipse Modeling Framework (EMF) models.
On the surface, it comes with the Viatra Query Language (\VQL) to formulate queries used to retrieve
information from modeling artifacts.
A \VQ query is essentially the same as a Datalog predicate.
VQL is similar to Datalog both in expressive power and in terms of its language constructs.
\VQ uses an incremental Datalog solver that can execute queries
incrementally in response to changes in the input model.
The model representation and change notification mechanism of \VQ is abstracted away from EMF, so
we can plug in any data representation that adheres to the \VQ APIs.

Under the hood, \VQ uses a computation network for incrementalization.
The input to the network is a set of relations.
A \VQ query is broken down into relational algebra operations (e.g.\ join, filter, union),
each having an incremental implementation.
Changes to the input must be communicated to \VQ in the form of tuple insertions and deletions.
Each node in the network caches its results.
In the face of incoming changes, the nodes incrementally update their results and then propagate
the \emph{delta} (that is, the difference between the old and the new result) on the outgoing edges.
The change propagation goes on until the result of each node stabilizes.
The fixpoint computation follows \SM just like in \CODEQL, but \VQ does
not need any auxiliary predicate to be present.

In terms of absolute performance of from-scratch execution, \CODEQL scales better and to larger
programs, but \VQ trades off initialization time and memory use for incremental update time.
For example, IncA~\cite{INCA_PLDI} which is an incremental static analysis framework based on \VQ could
achieve millisecond update times for inter-procedural analyses.
Compared to \CODEQL, \VQ does not employ so many optimizations at compile time, but \VQ is a
highly extensible framework, so optimizations or even new language features are fairly easy to add.
\VQ only uses dynamic join ordering based on simple heuristics.
\section{Incrementalizability of \CODEQL Analyses}
\label{sec:Impact}

This section investigates the potential for reuse when it comes to production \CODEQL analyses
and real-world codebases, thereby giving an estimate on how much we could gain with incrementality.

\begin{figure*}
\centering
\includegraphics[width=\textwidth]{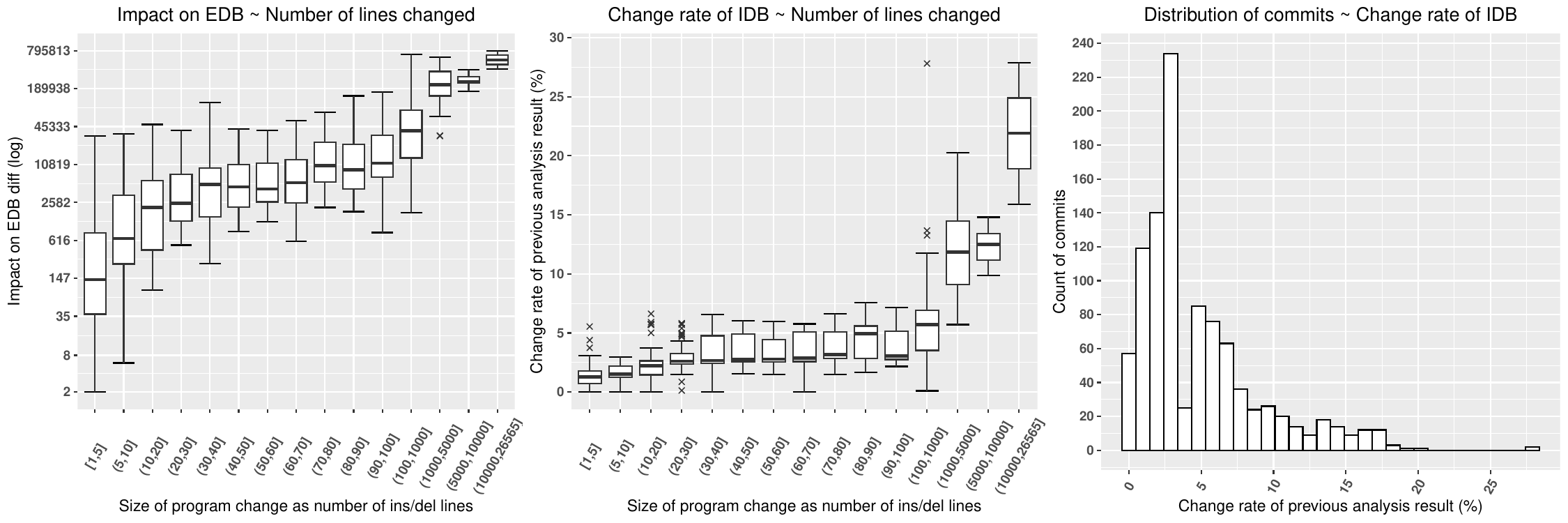}
\caption{Impact measurement results with optimized id generation on the spree project.}
\vspace{-5mm}
\label{fig:ImpactMeasurement}
\end{figure*}

\parhead{Measurement setup}
To measure the incrementalizability of \CODEQL analyses, we also perform impact measurements
inspired by prior work~\cite{INCA_PLDI,ELASTIC_SOUFFLE}.
We use real-world subject programs. The repositories we selected are all trending projects on GitHub with several thousand stars.
Our benchmark analysis is based on a production \CODEQL analysis used to find SQL injection
vulnerabilities.\footnote{\url{https://cwe.mitre.org/data/definitions/89.html}}
The original analysis is based on taint tracking across data flow paths, and it employs many of
the features that are responsible for good precision, e.g.\ inter-procedurality, context-sensitivity,
or field-sensitivity.
The analysis also has a lazy implementation, which means that it stops with the propagation of data flow facts as early as possible
if it finds that there are infeasible data flow paths.
Given that our benchmark codebases do not necessarily have SQL injection vulnerabilities, we
modified the original analysis by changing what kind of data flow nodes should be considered as
sources and sinks by the analysis.
We connect all formal parameters of the functions with all actual arguments of function calls to eliminate the laziness
of the analysis, thereby making it computationally expensive even in the absence of SQL injection vulnerabilities.
We selected altogether 6 Ruby projects for benchmarking. The complete list is in \autoref{sec:ImpactExt}.
We chose Ruby because the Ruby fact extractor has the simplest implementation out of all the
language front ends in \CODEQL.
It only requires parsing the source code, without any complex integration with a build system as
for compiled languages.
Choosing Ruby only really simplifies the fact extraction, but otherwise the analysis is
representative in terms of its complexity across the production \CODEQL analyses.

For each project, we use the newest 1000 commits from the history.
For each commit, we take the EDB, and we force \CODEQL to create an explicit representation of
the entire IDB (which it would not do otherwise due to optimizations).
We measure the size of each commit by summing up the number of lines of code that get deleted or
inserted.
A line modification is represented as a deletion-insertion pair.
We diff consecutive EDB and IDB snapshots to compute impacts.
We emphasize that there is no incrementality involved in these measurements.
We simply would like to understand first how impact relates to the commit size without an actual
incremental solver.

\parhead{Using the production \CODEQL pipeline} First, we used production \CODEQL for the
measurements. We report on the \emph{spree} project, which is an e-commerce platform
comprising $\sim$~73 KLoC Ruby code. We find that throughout the commit history the average
number of tuples in the EDB is $2.76$ million and the average number of tuples in the IDB is $217.41$
million. We find that \emph{irrespective of the commit size}, the average impact on the EDB
throughout the commit history is $4$ million tuples. This number is actually higher than the size
of an individual EDB. This means that, even for small commits, we essentially reconstruct a new EDB
because we delete most of the previous contents and then insert a large number of new
tuples. This is obviously not compatible with incrementalization. We would expect that, at
least for small commits, only a small fraction of the EDB changes. Unfortunately, we have
not even computed the analysis result: This is just for the representation of the subject program
in the EDB! The average impact on the IDB is $209$ million, which is again bad for incrementality.
For space reasons, we only talk about one specific project here, but, for all other projects we
considered, we found similar high impacts.

\parhead{Using stable ids} We investigated in detail the root cause of the high impacts,
and we found that to be due to the lack of stable ids for AST nodes and ADT values. Recall that the
trap importer rolls out ids sequentially based on a counter, so a small change to one file may
end up shifting ids for all the AST nodes later in the file \emph{and} all subsequently imported
files. Similarly, the \CODEQL solver assigns ids to ADT instances sequentially, again offering
no guarantee that the same ADT instance will be given the same id across different runs. This
leads to a huge number of spurious changes. We mitigate this problem in two steps.

First, we modified both the trap extractor and importer. In the extractor, we implemented a new
approach for constructing global keys for every AST node based on paths in the AST\@.
When we traverse the AST to create the trap files, we recursively assemble
a node path that describes how we can get to a node across containment edges. For
example, a path like $r\_1\_3$ gets associated with a node which can be reached from the root if
we take the first child under the root, then the third child after that. If we prepend the full
file path of a source file before such node paths, we get globally unique keys. We do
not use an id counter anymore in the trap importer. Instead, we calculate an integer hash from
the global key and use that in the actual tuples. This approach of course does not help in all
situations to ensure stability because e.g.\ it does not survive a node shifting across commits, but
it still significantly helps in id reuse. This kind of id generation is a must if we want to
incrementalize the front end of \CODEQL (cf. \chref{ch:frontend}).

Second, we employ a workaround to reuse ids for ADT instances. \CODEQL can dump the id
assignment of ADT instances to disk. As we proceed along commit pairs in the measurement, we dump
the id pool when we process the old commit, and we augment the analysis run for the new commit to
reuse the same id assignment whenever possible based on the pool.

Using these two improvements, \autoref{fig:ImpactMeasurement} shows our new measurements on the
spree project. \autoref{sec:ImpactExt} presents the results for several other projects. The first
sub-figure shows the correlation between the impact on EDB and the size of commit. The commits
are arranged into buckets in terms of how many lines of code they affect. The impact on the
EDB is proportional to the size of the commit, and the impact maxes out at around 800,000 tuples
which is much smaller than the overall size of the EDB. We show change rates for the IDB. The
change rate tells us the ratio of overall number of tuples deleted or inserted (to get to the new
IDB) to the size of the old IDB. The lower the value the better. The middle sub-figure
shows that the change rate is also proportional to the commit size. Up to 1000 changed lines in a
commit, the change rate is at or below 5~\%, which is exactly the kind of number we look for because
it demonstrates that we can gain a lot with incrementality. The right sub-figure
shows the distribution of the commits. The x-axis shows the change rate, and the y-axis shows how
many commits (out of the 1000) produce a certain change rate. The majority of the commits produce
a low change rate. The other codebases we considered also produced similar results.

\vspace{1mm}
\noindent \fbox{
\parbox{0.95\columnwidth}{
\noindent Regarding \chref{ch:incrementalizability}: Given that our benchmark analysis is
representative in terms of its complexity when it comes to production \CODEQL analyses and based
on the results of the impact measurements, we conclude that \CODEQL analyses are amenable to
incrementalization.

\noindent Regarding \chref{ch:frontend}: We found that the production \CODEQL front end is not
compatible with incrementalization. For an incremental analysis pipeline, it is crucial to employ
an id generation strategy that ensures stability as we process code changes. Our approach based
on node paths satisfies this requirement.
}}
\vspace{1mm}

\noindent The impacts measured here are solver-independent, and they give a lower bound on the amount of
work that any solver would need to do to update the analysis result. Even though we used the \CODEQL
solver in our setup, ultimately the analysis specification and the schema of the database
determine the impact values.

We also conducted some initial experiments with the Java front end of \CODEQL.
The Java front end is much more involved in the sense that it extracts information beyond the AST\@.
It actually invokes the build process of the given project (e.g.\ Gradle or maven
build) and dumps information about e.g.\ type hierarchy or generics, as well.
Due to these differences, it turned out to be technically much more difficult to employ our node path-based
id generation strategy because that would have required changes to the Java compiler.
Instead, we came up with a different strategy that ensures that the same ids get rolled out for
program elements originating from unchanged source files considering a commit pair without any
kind of id alignment guarantee for program elements originating from changed source files.
Due to space reasons, we discuss this strategy and present results only in \autoref{sec:ImpactExt}.
In sum, we found that the impact is proportional to the size of the code change in case of Java
codebases, as well.
\section{\IQL: Integrating \CODEQL with \VQ}
\label{sec:Backend}

As mentioned above, our prototype system \IQL works by translating \CODEQL queries to \VQ queries,
which are then run by the \VQ solver.
To translate from \QL to \VQ, we first run the \CODEQL compilation and optimization pipeline to
convert \QL (through DIL) to \CODEQL RA and, from there, compile to \VQ RA\@.
This has two main benefits: (i) RA is a much smaller language than \QL (or DIL), with
a lot of syntactic sugar removed, and (ii) we can take advantage of the optimizations that the
\CODEQL compiler performs.
\autoref{tab:Features} shows the main challenges of the translation and how we address them.

\begin{table*}
    \centering
    \begin{tabular}{l|ll|l}
        \toprule
        \textbf{Feature} & \textbf{\QL} & \textbf{\VQ} & \textbf{Translation} \\
        \midrule
        \SM evaluation & explicit & implicit & turn off auxiliary predicate generation \\
        join ordering & compile-time & runtime & modify \VQ to inherit the join order computed by \CODEQL \\
        \midrule
        tuple numbering & supported & not supported & add support to \VQ \\
        aggregation operators & many & few & add support for missing operators to \VQ \\
        built-ins & many & few & add support to \VQ by wrapping \CODEQL implementation \\
        parity stratification & supported & not supported & eliminate by rewriting \\
        recursive aggregates & \multicolumn{2}{c|}{different semantics} & eliminate by rewriting \\
    \end{tabular}
    \caption{Summary of how the translation from \QL to \VQ handles missing and divergent features}
    \vspace{-6mm}
    \label{tab:Features}
\end{table*}

\parhead{Challenges due to differences arising at runtime}
The first two challenges have to do with the way the systems deal with recursion and join
ordering. As mentioned in \autoref{sec:Problem-CodeQL}, \CODEQL implements \SM evaluation
by explicit code generation, while \VQ implements it natively in the solver. This discrepancy can
be resolved by turning off the pass of the \CODEQL compiler that generates the auxiliary
predicates.\footnote{Note
that this means the resulting program is no longer runnable by CodeQL, but this is
not a problem for our study.} Join ordering in \CODEQL is done at compile time, with
the RA statically encoding the join order. \VQ dynamically reorders joins at runtime. We deal
with this by extending \VQ to allow passing join-ordering hints taken from the \CODEQL compiler,
which the \VQ solver uses at runtime.

\parhead{Challenges due to different expressive power}
The remaining challenges have to do with missing or divergent features in \VQ.
Two commonly-used key features of \QL without a direct equivalent were
reimplemented from scratch. One is the so-called \emph{tuple numbering}
operator, which assigns a unique integer to each tuple in a relation and serves
as the basis of \CODEQL's implementation of ADTs. \VQ
does not have a direct equivalent and the \CODEQL implementation of this
operation is too deeply tied to the internals of \CODEQL to port, so
we instead rewrote it from scratch, using a simple global cache keeping a
mapping from all previously seen tuples to their unique identifiers. This is not
a plausible solution for real-world usage since we can never know when we do not
need a tuple anymore and hence the cache grows without bound, but it suffices
for our study. The other key feature we reimplemented are \QL's aggregation operations on
strings that have no direct \VQ equivalent, but were fairly straightforward to
add.

\CODEQL's built-in operations are a bit more complicated. While \VQ offers similar operations in
some cases (e.g.\ transitive closure), we decided against adapting them or reimplementing missing
built-ins from scratch to avoid the risk of introducing subtle incompatibilities. Instead, we
directly lifted the Java implementation of these built-ins from the \CODEQL engine, wrapped in a
translation layer to convert between the respective storage formats for \CODEQL and \VQ relations.
This ensures not only that the operations are semantically equivalent, but also that they
have the same performance. However, the \CODEQL built-ins are not incremental, and fitting them
into the otherwise incremental evaluation in \VQ is somewhat involved. There is a \VQ computation
node behind every built-in appearing in the analysis. Whenever such a node gets notified about
changes to its input relations, it computes the updated input relations in full, and feeds them to the
wrapped \CODEQL built-in (which can only work on complete relations, not on deltas). The result
of the built-in is then diffed against the previous result, and the delta is propagated to the
dependent nodes. Given that \VQ normally propagates individual tuples without batching them,
repeatedly re-evaluating a built-in from scratch can easily become a performance bottleneck. To
avoid this situation, we modified the implementation of the computation network in \VQ to (i)
batch the inputs of built-ins and (ii) schedule the invocation of nodes evaluating built-ins only
after all their incoming updates have arrived.

The two most complicated features of \QL that we had to deal with are parity-stratified
recursion and recursive aggregates. However, it turns out that in practice neither of these two
features is very widely used, and the code that does use them can be rewritten to avoid them, at
some cost in readability. Since our aim is to study the potential for incrementality in \CODEQL,
not to provide a complete incremental \CODEQL implementation, we decided to take this approach.

\parhead{Hybrid evaluation}
While the translation outlined before works and produces correct \VQ programs whose results can
be updated very efficiently, these programs are much bigger than what \VQ has been designed to
handle. As we will see in \autoref{sec:Evaluation}, even on small subject programs, the memory
use of \IQL can be prohibitive.

To address this problem, we study a \emph{hybrid} evaluation strategy where parts of the
program are evaluated non-incrementally by \CODEQL, and the results are integrated into the
overall \VQ program similar to how built-ins are evaluated. The reason why this is interesting is
that the parts of the analysis that are evaluated non-incrementally do not require intermediate
caching but only at the point where the final (partial) results appear. By choosing which parts
of the program to evaluate non-incrementally, we can trade off \CODEQL's superior scalability
for \VQ's incrementality.

\begin{figure}
    \centering
    \includegraphics[width=0.7\columnwidth]{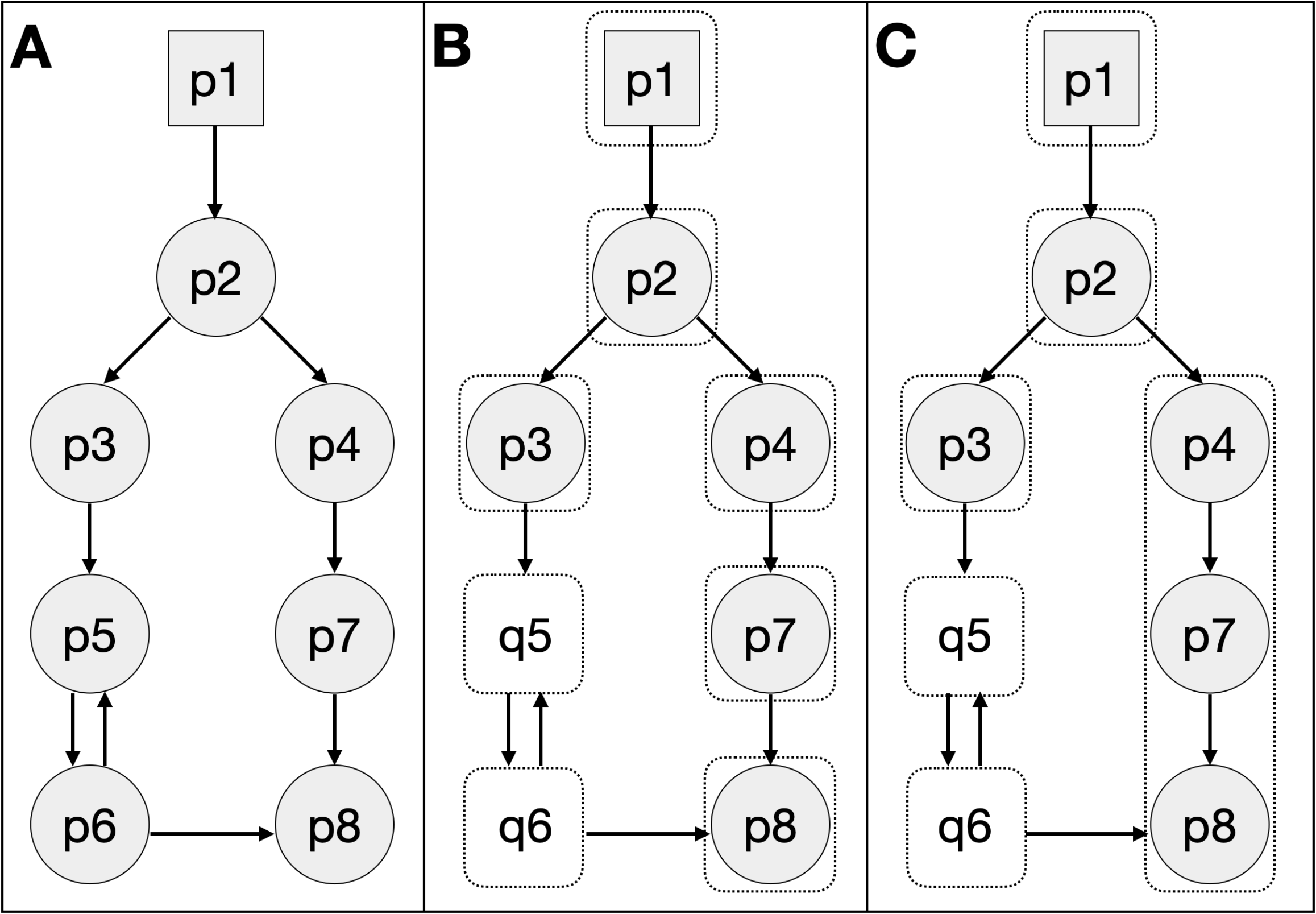}
    \caption{Transformation strategies from \CODEQL to \VQ: (A) Example dependencies between
    \QL predicates, (B) transformation strategy which creates one \VQ query per \QL predicate,
        (C) chaining transformation strategy.}
    \vspace{-5mm}
    \label{fig:Hybrid}
\end{figure}

As mentioned above, we have to turn off \CODEQL's \SM transformation to enable translation
to \VQ. This means that the recursive predicates have to be evaluated by \VQ. We still have
freedom in deciding how we evaluate the non-recursive predicates.
Consider the \QL program shown schematically in \autoref{fig:Hybrid}~A: It consists of one EDB
predicate \ic{p1}, and seven IDB predicates \ic{p2} through \ic{p8}. The arrows indicate
dependencies, so, for example, predicate \ic{p2} depends on \ic{p1}, while \ic{p5} and \ic{p6}
recursively depend on each other.
For the non-recursive predicates, a very simple approach would be to wrap each
of them individually in a \VQ query that calls out to \CODEQL for evaluation, translating back
and forth between the respective storage formats similar to how built-ins are handled.
\autoref{fig:Hybrid}~B shows this approach: each of \ic{p1}, \ic{p2}, \ic{p3}, \ic{p4}, \ic{p7},
and \ic{p8} are wrapped in a \VQ query, while \ic{p5} and \ic{p6} are translated into \VQ queries
\ic{q5} and \ic{q6}. This approach, however, makes things worse: Not only are the wrapped
predicates evaluated non-incrementally, but the transformation happening at the \VQ-\CODEQL
boundary introduces significant overhead both in memory use and run time.

To address this, we use a refined translation strategy shown in \autoref{fig:Hybrid}~C\@. As
before, EDB predicates are individually wrapped in \VQ queries. IDB predicates are grouped into
\emph{chains}, where a chain is a maximal sequence of (non-recursive) IDB predicates that each flow
into at most one other predicate. In the example, we identify one non-trivial chain \ic{p4, p7, p8},
while the remaining predicates form a single-element chain each. Note that while it may
sound like a good idea to group all the non-recursive predicates together, we can't actually
do that because they may interact with recursive predicates. Also, the requirement for at most one
outgoing flow is intentional because this way, if a predicate is a fork point (like \ic{p2}), then
we will benefit from the caches (at \ic{p2}) shared by the fork endpoints (\ic{p3} and \ic{p4}).
In the following, when we use the term \emph{hybrid} solver, we refer to the approach based on chaining.
\vspace{-2mm}
\section{Performance Benchmarking of \IQL}
\label{sec:Evaluation}

This section investigates all research challenges formulated in \autoref{sec:Problem} that
are related to performance.

\parhead{Measurement setup}
We benchmark the performance of \IQL with real-world Ruby projects and their commit histories using
the same benchmark analysis that we used for the impact measurements in \autoref{sec:Impact}.
Technically, the Ruby front end is implemented in Rust, and \IQL and \VQ are implemented in Java.
\IQL is a closed-source project because it integrates tightly with \CODEQL, and \CODEQL is a
closed-source project.
Due to scalability issues in memory use, we only use two smaller Ruby projects:
\href{https://github.com/ddnexus/pagy}{\color{blue}{pagy}} comprising 6~KLoC code and
\href{https://github.com/errbit/errbit}{\color{blue}{errbit}} comprising 9~KLoC code.
For each project, we use 200 commits from the history.
Our Ruby front end makes use of the id generation strategy described in \autoref{sec:Impact}, but
the front end itself is not incremental.
Instead, we precompute the EDB differences: Given every commit pair $(c_{old}, c_{new})$, we
compute $EDB_{old}$ and $EDB_{new}$ and then diff them.
We then perform a from-scratch analysis using $EDB_{old}$ to compute $IDB_{old}$.
We make use of both kinds of incremental solvers in \IQL: the \emph{fully-incremental} one which
only executes the built-ins non-incrementally and the \emph{hybrid} one which makes use of
production \CODEQL for the execution of non-recursive predicates, as well.
We measure the wall-clock initialization time.
Then, we take the EDB delta between $EDB_{old}$ and $EDB_{new}$ and let \IQL perform an incremental
update on $IDB_{old}$ to compute $IDB_{new}$.
We measure the incremental update time and also the stationary memory use (after explicit GC
invocation) of \IQL before and after the incremental update.
We automated all of these steps and executed the benchmarks on GitHub Actions.

\parhead{Correctness}
To ensure that \IQL performs correct incremental updates, we have set up automated
verification machinery.
Before and after an incremental update, we check that \IQL computes the exact same results as
what production \CODEQL would compute.
We diff all relations in the IDBs, and we make sure that those are identical.
The only difference we allow is the actual values of fresh ids, but that is expected as the two
systems use different strategies.

\begin{figure}
\centering
\includegraphics[width=\columnwidth]{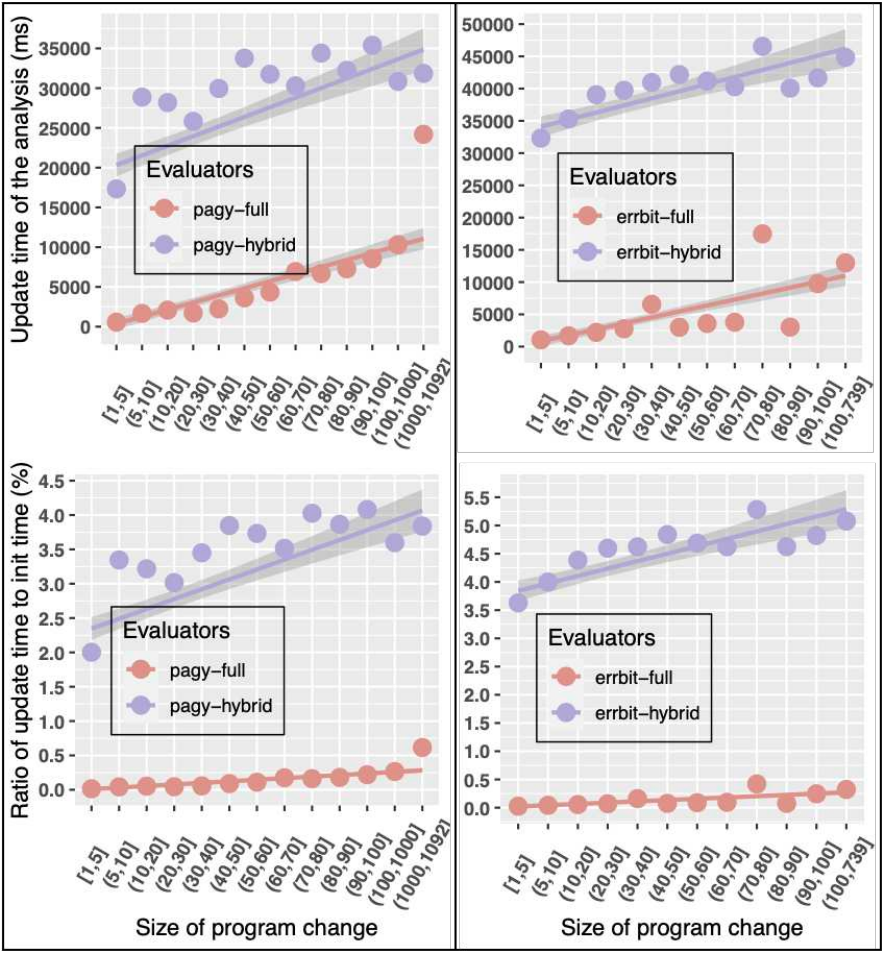}
\caption{Results of the performance benchmarking: pagy on the left and errbit on the right.}
\vspace{-5mm}
\label{fig:Benchmark}
\end{figure}

\parhead{Results}
\autoref{fig:Benchmark} shows our benchmark results.
The plots on the top show the incremental update times relative to the size of the code
changes.
The commits are grouped into change size buckets.
The dots represent the mean value in a given bucket.
The gray areas show the standard deviation of the values with 95~\% confidence.
The lines are shown to provide visual guidance, and they are computed as a linear regression fit
onto the data points.
We show the results for both codebases using the two evaluator approaches.
The plots at the bottom show the ratio of the incremental update times relative to the
initialization time.
We summarize the initialization time and memory use below:
\begin{center}
\begin{tabular}{lrrrr}
\toprule
& \multicolumn{2}{c}{\emph{full}} & \multicolumn{2}{c}{\emph{hybrid}} \\[1ex]
& init (min) & mem (GB) & init (min) & mem (GB) \\
\midrule
\emph{pagy}   & $66.10$ & $70.68$ & $14.40$ & $21.52$ \\
\emph{errbit} & $67.10$ & $72.61$ & $14.72$ & $22.51$ \\
\bottomrule
\end{tabular}
\end{center}

\noindent In terms of the incremental performance, we see very promising results.
There is a linear correlation between the update time and the commit size in all cases.
The fully incremental approach is better in terms of update times than the hybrid one, but this
is expected, as the latter one executes a large part of the analysis non-incrementally.
For all size buckets, the fully incremental approach delivers update times in around 15 seconds
or less, while the hybrid one is also well within the one-minute ballpark.
The ratio of update time to initialization time is less than 1~\% in case of the fully incremental
approach, which shows that \IQL can indeed exploit the potential for speed-ups as forecast in
\autoref{sec:Impact}.
While it is not a fair comparison, it is worth pointing out that the production \CODEQL
solver requires around a minute to execute the same analysis from scratch on these codebases when
all caches are turned off.\footnote{Turning off all caches does make sense because a realistic
scenario is to use \CODEQL on GitHub Actions after a push to a branch in a new virtual machine,
so caches would not be readily available anyway.}
This means that the fully incremental approach is actually competitive in this regard against a
production system, even though it is only a prototype.

The initialization time is now a different story when it comes to the fully incremental approach.
We see run times of about an hour, which is significantly slower than production
\CODEQL.
However, the hybrid approach presents an interesting trade-off opportunity, as its initialization
time is only around 15 minutes.
Considering that the incremental update times of the hybrid one were also less than a minute, this option may
present a good choice in general.
Interestingly, the initialization times do not change much between the two codebases.
It turns out that our benchmark analysis produces a very complex and large computation network in
\VQ, and building that actually becomes a dominating factor in the run time.
These differences also make it clear that production \CODEQL simply scales significantly better
than \VQ.
We argue though that even the one-hour initialization time is acceptable, as we could precompute
the results for select ``main'' branches and compute the results for a PR with incremental
updates from those.

The biggest obstacle right now is the memory use.
The several tens of GB values for the fully incremental approach are prohibitive.
The hybrid approach helps to reduce these numbers significantly, but even \textasciitilde~20
GB is too large for a regular GitHub Actions runner.
Even with these results, we believe that our study is useful because it demonstrates the
potential of incrementalization.

\vspace{1mm}
\noindent \fbox{
\parbox{0.95\columnwidth}{
\noindent Regarding \chref{ch:updateTime}:
The incremental update times are within the 1 minute ballpark for both approaches, and the
incremental update time is proportional to the commit size.

\noindent Regarding \chref{ch:initTime}:
The fully-incremental approach requires around 1 hour to perform a from-scratch run, while the
hybrid approach requires around 15 minutes. Both of these values are acceptable given that they
are not too high in absolute terms, and we can precompute results.

\noindent Regarding \chref{ch:memory}:
The memory use of the fully-incremental approach is very high, but the hybrid approach comes
with significantly better memory use. More work would be needed to improve the memory use in a
production incremental solver.
}
}
\section{Related Work}
\label{sec:Related}

In \autoref{sec:Problem-Incrementalizability}, we already discussed prior work on the
incrementalizability of Datalog-based static analyses. Here, we contrast our work to other
related areas.

\parhead{Studies on the use of static analyses} There are several studies in the literature which
investigate what developers expect from static analyses.
We highlight some key observations here.
\citeauthor{WHAT_DEVS_WANT_AND_NEED} point out that (i) developers in their study care a lot about
low false positive rate, (ii) 77~\% of them would appreciate if an analysis could be governed to
only analyse a change as opposed to the entire codebase, and (iii) long-running analyzers
that exceed a few minutes would not be considered by nearly three quarters of
developers in their study~\cite{WHAT_DEVS_WANT_AND_NEED}.
\citeauthor{WHY_DONT_DEVS_USE_ANALYSES} explain that developers expect static
analyses to be properly integrated into their workflows, whether that is the IDE or PR reviews~\cite{WHY_DONT_DEVS_USE_ANALYSES}.
Moreover, it is important that feedback comes in a timely manner before developers switch
context because that makes the interpretation of the analysis output more difficult.
\citeauthor{TRICORDER} investigate the use of static analyses at Google~\cite{TRICORDER}.
They find similar observations as the previous studies in terms of expected run time and workflow
integration.
However, they also provide the actual list of static analyses that they use at scale.
It turns out that all the analyzers perform some sort of linting or dependency analysis.
These are obviously helpful analyses, but they are computationally not too expensive, and they
immediately ``incrementalize'' on a file-level granularity.
In contrast, our goal is to deliver fast update times for sophisticated data flow analyses that
are computationally much more expensive.

\parhead{Incrementalizing individual static analyses} Incrementalization has received a lot of
attention both as a means to speed up specific static analyses or as a technique to power entire
analysis frameworks dealing with classes of analyses.
For example, \citeauthor{POINTS_TO_CFLR} devise an incremental points-to analysis by
formulating it as a graph reachability problem and then making use of well-known techniques for
incrementalizing graph reachability itself~\cite{POINTS_TO_CFLR}.
\citeauthor{INC_DEMAND_POINTS_TO} design an incremental and demand-driven points-to analysis
formulated in Datalog~\cite{INC_DEMAND_POINTS_TO}.
Demand-driven means that the analysis only computes points-to information that is relevant to the
client.
They achieve this with the use of magic set transformation~\cite{MAGIC_SETS} which essentially introduces
auxiliary predicates in the Datalog program encoding the demand itself (e.g.\ fixing the pointer
variable to something specific).
The incremental maintenance in their solution is based on Delete and Rederive
(DRed)~\cite{MAINTAINING_VIEWS_INCREMENTALLY} which is a well-known algorithm for incrementalizing recursive
Datalog programs.
While \IQL is an analysis \emph{framework} that does not make assumptions about the kind of \QL
analysis it needs to incrementalize, there are definitely connections to these pieces of related
work.
Magic set transformation is heavily used in the \CODEQL compiler to reduce work by inlining
information about the input program, so \IQL benefits from that, too.
DRed is also responsible for the incrementalization in \IQL because that is one of the
incremental solver algorithms used in \VQ.

\parhead{Incremental static analysis frameworks}
There are several approaches that incrementalize set-based static analyses~\cite{POLLOCK_SOFFA_INCDF,KhedkerPHD,MAGELLAN_FRAMEWORK}.
Set-based here refers to how the analyses operate: They propagate sets of data-flow facts along
the nodes in the control flow graph using set union or intersection at merge points.
Many practically interesting static analyses can be formulated this way (such as liveness or
uninitialized read), but this expressive power is not enough for production \CODEQL analyses.

The Reviser framework incrementalizes IFDS/IDE analyses~\cite{REVISER}.
IFDS/IDE is a generic framework for formulating data flow analyses, the former supporting only
power set lattice values as data flow facts, while the latter supporting custom lattice values.
IFDS/IDE boil down to graph reachability and summarization of the effects of
functions.
While IFDS/IDE have been used to implement industry-strength static analyses (e.g.\ Boomerang~\cite{BOOMERANG}), the
challenge in efficiently employing the technique is that summaries should encode function-local
information only, and this is difficult to achieve in practice~\cite{BODDEN_SECRET_SAUCE}.
It is challenging to compare the expressive power to that of \QL because of the different
computational models.

Infer is a static analysis framework that is used at scale at Facebook~\cite{FACEBOOK_INFER}.
Infer supports analyses that reason about manipulations of the heap (e.g.\ null pointer
dereference, finding resource leaks).
The theoretical foundation is separation logic which allows for efficient summarization of
effects of heap manipulations.
This approach lends itself to efficient incrementalization because summaries only need to be
re-computed for changed code parts, and then the overall analysis result can be composed of
the individual summaries.

\INCA~\cite{INCA_ASE, INCA_PLDI} is an incremental static analysis framework that also builds on \VQ.
IncA is unique among incremental Datalog-based systems in that it supports user-defined lattices
and recursive aggregations over lattices which is an important building block for static analyses.
Similar to \CODEQL, IncA also defines a higher-level DSL for analysis specification which then
gets compiled to the RA format of \VQ.
Given that \QL does not support custom lattices, we also did not make use of this feature in \VQ.

\parhead{Compilation of other languages to Datalog}
There are several pieces of related work that revolve around the idea that while Datalog has many
benefits when it comes to the execution aspect, perhaps it does not have the best design when it
comes to the specification aspect.
\QL itself is a good example because it adds an object-oriented flavor to Datalog with higher-order
domain-specific extensions.
\citeauthor{SYSTEMATIC_INC_TYPE_CHECKERS} design a DSL for defining type checking rules and then
compile the DSL code to Datalog to be executed by \INCA~\cite{SYSTEMATIC_INC_TYPE_CHECKERS}.
The challenge in their work is that Datalog solvers expect the input relations to be finite, but, when
it comes to type checking, the typing relation itself can grow infinite.
They make heavy use of magic set transformations to get rid of the typing context in the typing
relation, thereby eliminating the problem of infinite input.
\citeauthor{FUNCTIONAL_PROGRAMMING_WITH_DATALOG} later go a step further by designing Functional \INCA
which is still a limited but general-purpose functional language that also compiles to \INCA
Datalog~\cite{FUNCTIONAL_PROGRAMMING_WITH_DATALOG}.
Datafun~\cite{DATAFUN} is not strictly speaking an example where the language itself gets compiled to
Datalog, but it is a language which marries functional programming and Datalog concepts.
Datafun comes with an explicit fixpoint operator that can be used to mark specific functions
where fixpoint computation shall happen.
This is in contrast to \QL or Functional \INCA where fixpoint computation is ultimately
determined by the dependencies between predicates/functions.

\section{Conclusions}

This study successfully showed that incrementalization has potential when it comes to speeding up
production \CODEQL analyses.
First, we showed this empirically with our impact measurements, and then we also demonstrated
this by creating our prototype incremental solver.
\IQL delivers updated feedback in sub-minute run time for commit-style changes.
The price of the fast update time is the high initialization time and memory use.
While the initialization time is acceptable, the memory use requires further work before
incrementalization can make its way to production \CODEQL.
\section*{Acknowledgement}

Thanks to André Pacak, Ian Wright, Gábor Bergmann, Matthias Plappert, Max Schäfer, Nathan Corbyn,
Oege de Moor, Rahul Pandita, and the members of the \CODEQL Core team for helpful feedback and
discussions on this work.

\balance

\clearpage

\bibliographystyle{ACM-Reference-Format}
\bibliography{main}

\clearpage
\appendix
\section{Follow-up on Incrementalizability of \CODEQL Analyses}
\label{sec:ImpactExt}

\parhead{Impact measurements on Ruby codebases}
In \autoref{sec:Impact}, we only provided details about a single Ruby codebase called spree.
Here we share the results for several other projects.
\autoref{tab:Projects} summarizes the projects and some analysis-specific metrics about them.
For each project, we used 1000 commits from the history.

\parhead{Impact measurements on Java codebases}
In \autoref{sec:Impact}, we briefly talked about our results on Java codebases. Here, we
elaborate further on our approach and provide some detailed results.
As mentioned before, it was technically much more challenging to implement the node path-based id
generation strategy for Java because the fact extractor also dumps derived information about the
subject programs which requires compilation and build tasks to be executed.
Instead, we implemented a simpler approach to ensure id stability on the file level across the
two sides of a commit pair.
We modified the trap extractor to emit a \ic{bump\_id\_counter} directive at the end of every trap
file.
When the trap importer sees this directive, it increments the global id counter to the next
million value.
The purpose of this is to ensure, with high probability, that the id counters used in the two
sides of a commit pair will ``line up'' again even if a trap file belongs to a changed source file.
We have enough headroom for bumping to the next million because \CODEQL uses 64-bit integer ids, so
the value range is large enough.
While this approach would not work in general because of potential overflows, the Java projects
we considered did not have that many source files that they would have exhausted the value range.
We would have caught overflow issues anyway because the \CODEQL solver would throw a runtime
error in that case.

Using the id-bumping strategy, we present the results for two further Java projects in \autoref{tab:Projects}.
For the Java codebases, we used 500 commits and a \CODEQL analysis that
checks for improper validation of array indices (CWE-129).\footnote{\href{https://cwe.mitre.org/data/definitions/129.html}{https://cwe.mitre.org/data/definitions/129.html}}

\onecolumn

\begin{tabularx}{0.98\textwidth}{llXr|rrrr}
   \toprule
  & name / plot & short description & size & avg. EDB size & avg. EDB impact
  & avg. IDB size & avg. IDB impact \\
  &&& (\emph{KLoC}) & \multicolumn{4}{c}{\emph{all values are in million tuples}} \\
  \midrule

  \multirow{6}{*}[-30pt]{\rotatebox{90}{Ruby}} &
  \href{https://github.com/ddnexus/pagy}{\color{black}{pagy}} / \autoref{fig:Pagy-Full} &
  pagination library for Ruby & 6 & 0.18 & 0.01 & 13.07 & 0.85 \\

  & \href{https://github.com/errbit/errbit}{\color{black}{errbit}} / \autoref{fig:Errbit-Full} & a tool
  for collecting and managing errors from other applications & 9 & 0.28 & 0.01 & 22.46 & 1.31 \\

  & \href{https://github.com/backup/backup}{\color{black}{backup}} / \autoref{fig:Backup-Full} &
  a system utility that makes it easy to perform backup operations & 24 & 0.43 & 0.02 & 31.97 & 1.63 \\

  & \href{https://github.com/diaspora/diaspora}{\color{black}{diaspora}} / \autoref{fig:Diaspora-Full} &
  a privacy-aware, distributed, open-source social network & 52 & 1.67 & 0.15 & 129.62 & 12.82 \\

  & \href{https://github.com/spree/spree}{\color{black}{spree}} / \autoref{fig:ImpactMeasurement} & an
  e-commerce platform & 73 & 2.76 & 0.05 & 217.41 & 9.68 \\

  & \href{https://github.com/fastlane/fastlane}{\color{black}{fastlane}} / \autoref{fig:Fastlane-Full} &
  an automation tool for building and releasing iOS and Android apps & 118 & 3.65 & 0.01 & 278.96 & 7.84 \\

  \midrule

  \multirow{2}{*}[-10pt]{\rotatebox{90}{Java}} &
  \href{https://github.com/Netflix/zuul}{\color{black}{zuul}} / \autoref{fig:Zuul-Full} &
  Netflix gateway service implementation & 25 & 0.69 & 0.08 & 37.68 & 4.95 \\

  & \href{https://github.com/apache/opennlp}{\color{black}{opennlp}} / \autoref{fig:OpenNLP-Full} &
  Natural language processing library & 79 & 1.52 & 0.15 & 59.7 & 2.36 \\

  \bottomrule
 \end{tabularx}
 \captionof{table}{Summary of projects used in the impact measurements.}
 \label{tab:Projects}

\newpage

\begin{figure}
 \centering
 \includegraphics[width=\textwidth]{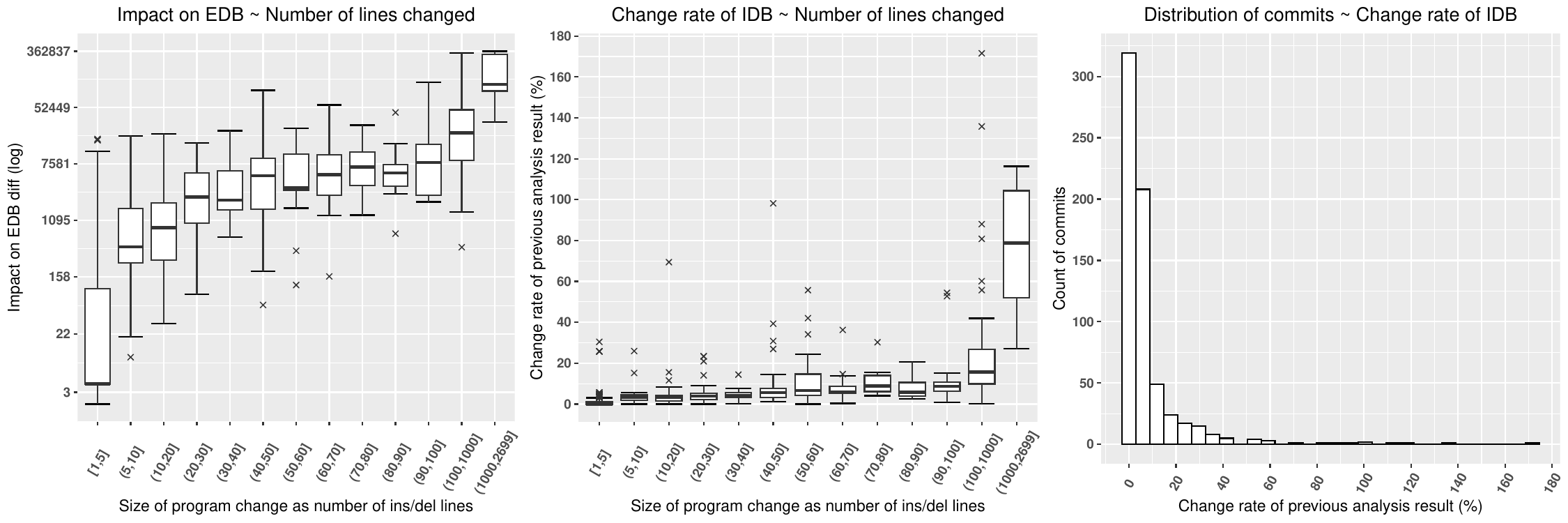}
 \caption{Results of impact measurements on pagy.}
 \label{fig:Pagy-Full}
\end{figure}

\begin{figure}
 \centering
 \includegraphics[width=\textwidth]{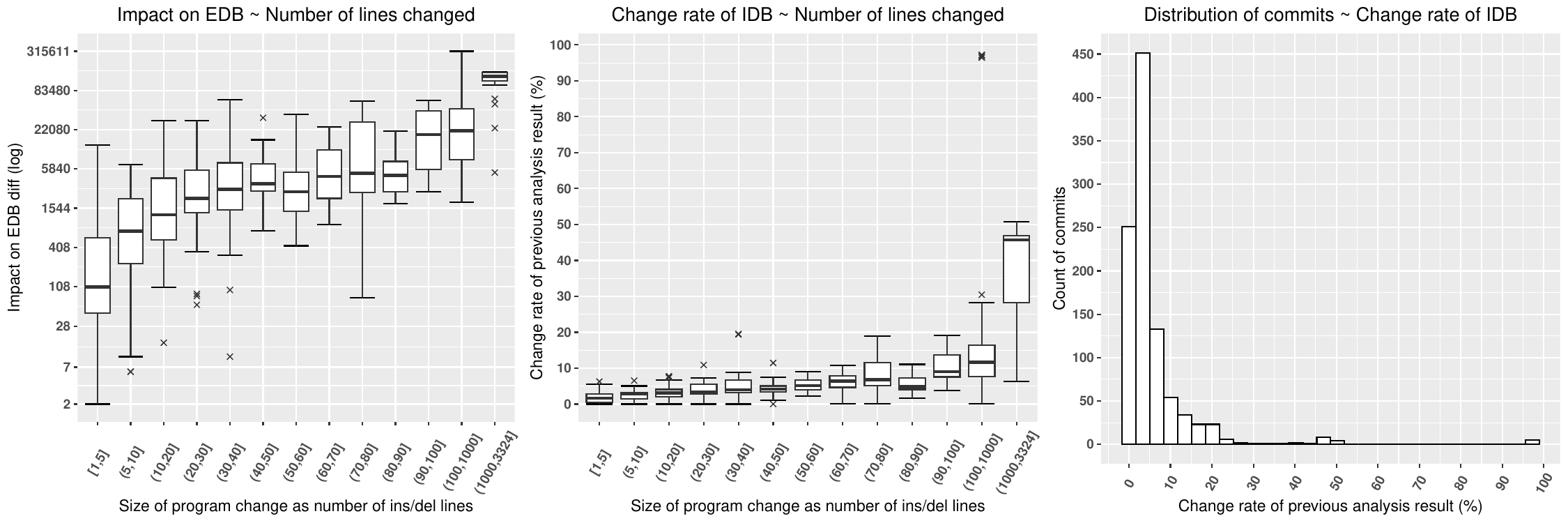}
 \caption{Results of impact measurements on errbit.}
 \label{fig:Errbit-Full}
\end{figure}

\begin{figure}
 \centering
 \includegraphics[width=\textwidth]{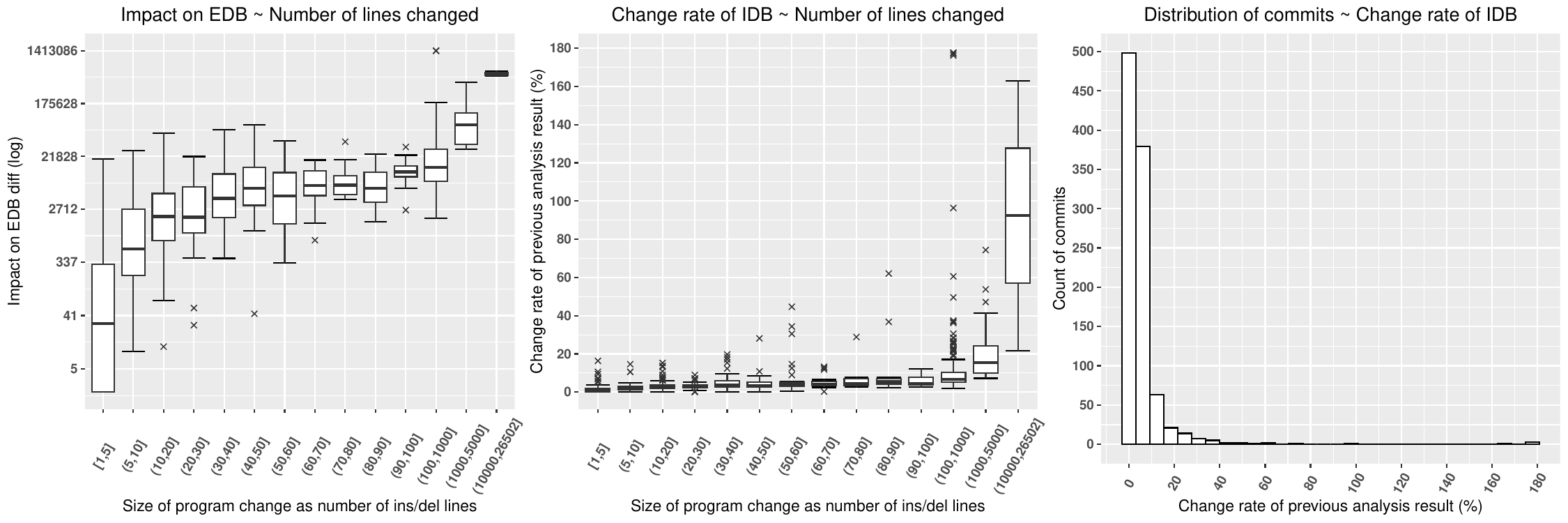}
 \caption{Results of impact measurements on backup.}
 \label{fig:Backup-Full}
\end{figure}

\begin{figure}
 \centering
 \includegraphics[width=\textwidth]{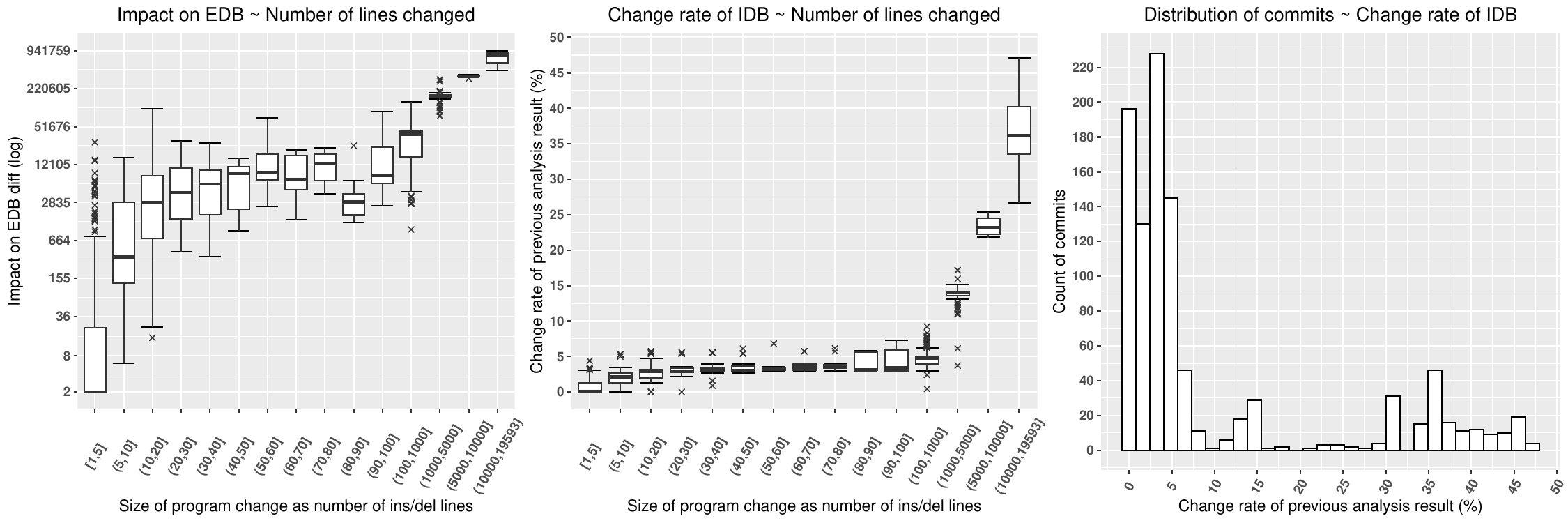}
 \caption{Results of impact measurements on diaspora.}
 \label{fig:Diaspora-Full}
\end{figure}

\begin{figure}
 \centering
 \includegraphics[width=\textwidth]{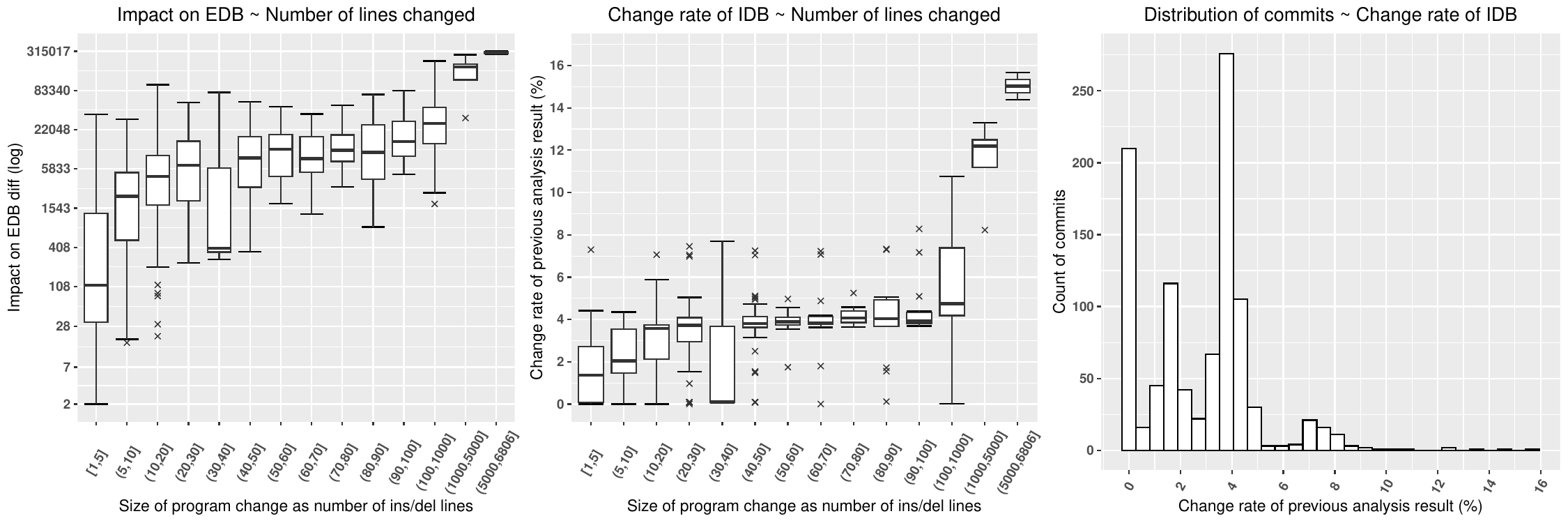}
 \caption{Results of impact measurements on fastlane.}
 \label{fig:Fastlane-Full}
\end{figure}

\begin{figure}
 \centering
 \includegraphics[width=\textwidth]{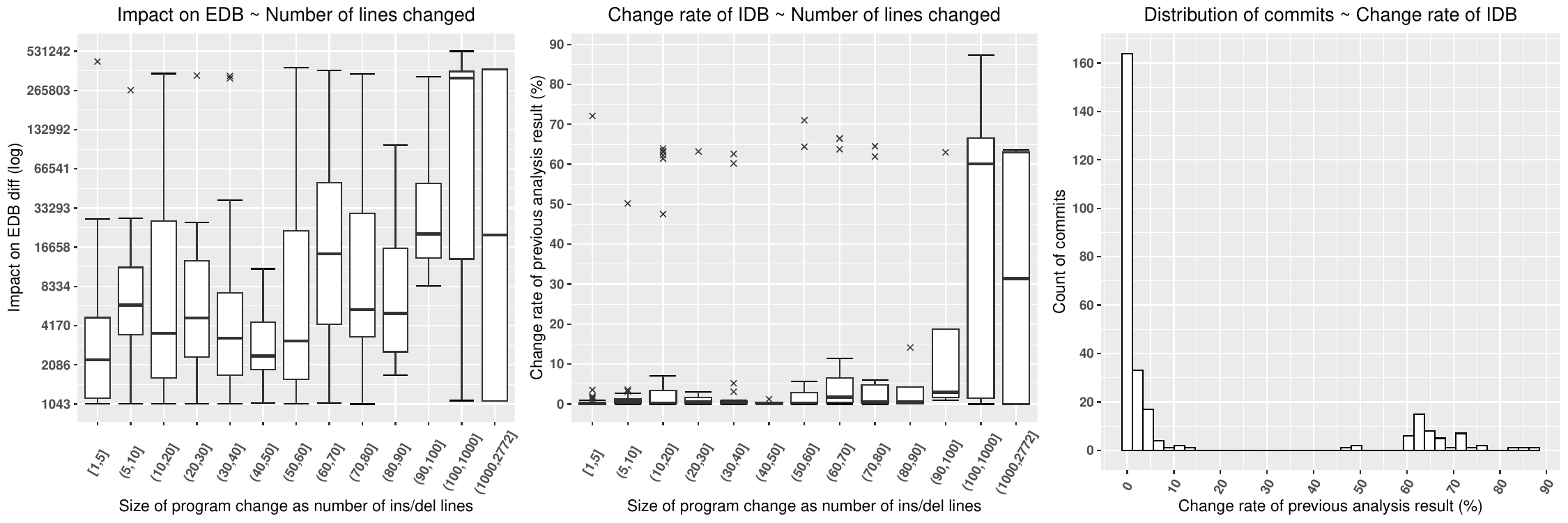}
 \caption{Results of impact measurements on zuul.}
 \label{fig:Zuul-Full}
\end{figure}

\begin{figure}
 \centering
 \includegraphics[width=\textwidth]{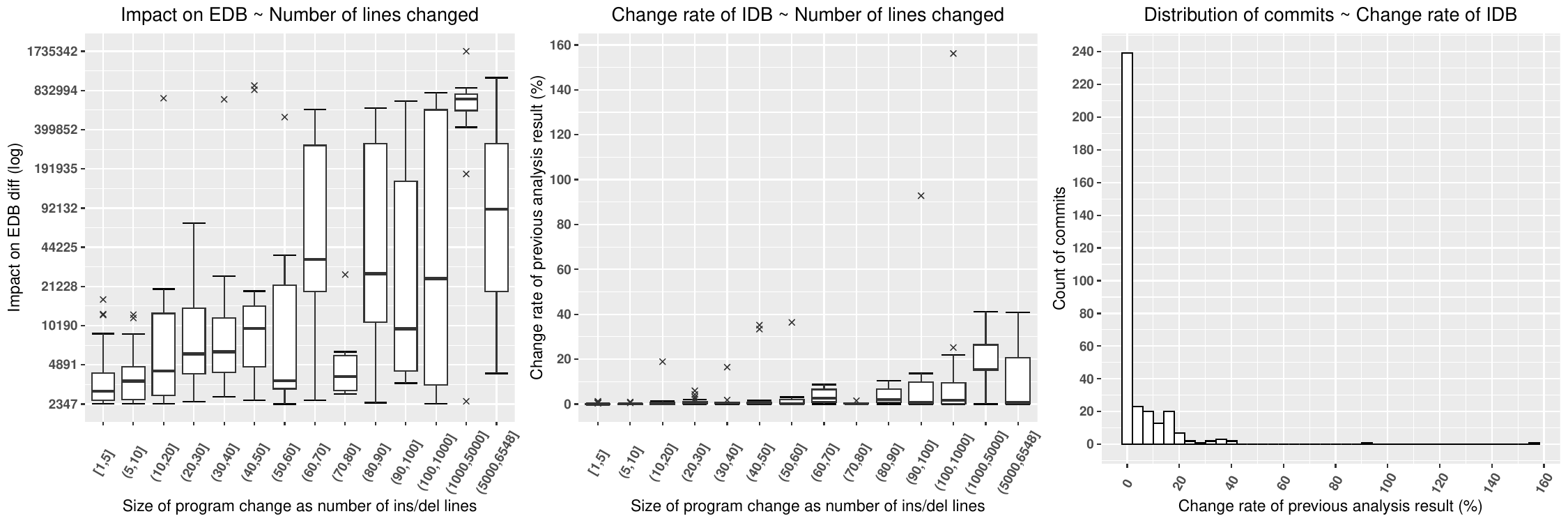}
 \caption{Results of impact measurements on opennlp.}
 \label{fig:OpenNLP-Full}
\end{figure}

\end{document}